\documentclass[12pt]{article}
\usepackage{blois,graphics}

\def\msun{{\rm\,M_\odot}}

\begin{document}

\bibliographystyle{plain}

\heading{Cosmological evolution and hierarchical galaxy formation}

\author{L. Miller \& W.J. Percival}
{Dept. of Physics, University of Oxford, Nuclear \& Astrophysics Laboratory,}
{Keble Road, Oxford OX1 3RH, U.K.}

\begin{bloisabstract}
\noindent
We provide a new multi-waveband compilation of the data describing the
cosmological evolution of quasars, and discuss a model that
attributes the evolution to variation in the rate of
merging between dark halos in a hierarchical universe.  We present a new
Press-Schechter calculation of the expected merger rate and show that
this can reproduce the principal features of the evolution.
We also show that the evolution
in the star-formation history of the universe is well-described
by this model. 
\end{bloisabstract}

\section{Cosmological evolution of quasars}

The aim of this paper is to interpret the strong cosmological evolution
that is seen in the quasar population.  We have known for many years
that the quasar comoving space density increases by a factor about
100 between a redshift of 0 and 2.  There is now strong evidence that
$z \sim 2$ marks a peak in quasar space density$^{\cite{shaver,ssg}}$.  
Fig.~1 shows a new compilation$^{\cite{wjplm}}$ 
of quasar comoving space density in a variety of wavebands.  
The samples cover a fixed range of luminosity, and the luminosity range 
is that
for which the greatest amount of evolution is observed, thereby providing
the strongest test of any explanations of evolution. 
Samples in different wavebands inevitably
sample different subsets of the overall AGN population, and the
luminosity ranges chosen probably correspond to slightly different
bolometric luminosities.  Thus there are differences in the normalisation
of each sample which have been removed by fitting
a smooth function to all the data and allowing the normalisations of each 
sample to vary so as to minimise the scatter about the curve.  The
renormalised sample points have been plotted with the normalisation appropriate
to the X-ray data.  The cosmological parameters of the $\Lambda$CDM
model have been assumed: the shape of the observed evolution shows little
dependence on choice of cosmology.

\begin{figure}[!tb]
  \centering
  \resizebox{5in}{3in}{
    \includegraphics{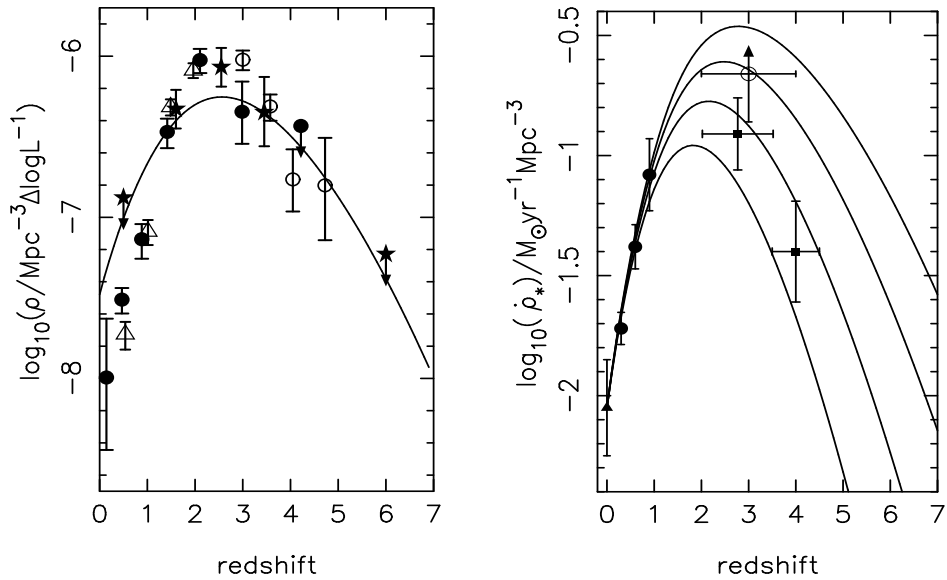}
  }
  \caption{\small {\em (Left:)}
  The comoving space density of flat-spectrum radio-loud 
  (\emph{stars}), UVX (\emph{open triangles}), Ly-$\alpha$ 
  (\emph{open circles}), and ROSAT 
  (\emph{filled circles}) selected quasars$^{\cite{wjplm}}$
  with the $\Lambda$CDM model for a 
  halo mass of $10^{11.8} \msun$. 
  {\em (Right:)} The mean comoving
  star formation rate$^{\cite{lilly,madau,pettini,hughes,gallego}}$
  ($\Omega_{\Lambda}=0$, $\Omega_M=1$, $h=0.5$) with 
  the $\Gamma$CDM model for masses of
  $10^{10.0} - 10^{11.5} \msun$ normalised to the
  local rate.  
} 
\end{figure}

The agreement between the different wavebands is 
a strong argument that the apparent turnover in space density
is real:  any selection effects should be
different in the different wavebands and such good agreement would not be
expected.  \cite{shaver} have argued strongly that the flat-spectrum
radio-loud quasars in particular are unlikely to be affected by any
selection- or non-evolutionary cosmological effects.  
We shall assume that the observed turnover is 
a real symptom of the cosmological evolution of the quasar population.

There are two chief hypotheses which have been invoked to
explain the cosmological evolution.  The first is based on
evolution in accretion rate onto black holes (e.g.
\cite{cavaliere}).  The other assumes that mergers  
trigger quasar activation and uses cosmological
models of galaxy formation to predict the observed evolution.
We are hampered by not knowing the lifetimes
of quasars, but there is growing evidence that the second picture is 
correct.  At least some quasars and active
galaxies are associated with mergers$^{\cite{stockton,canalizo}}$.  
There are 
also links between starburst galaxies and AGN, in that many 
ultra-luminous infrared galaxies appear to harbour obscured AGN, and those
ULIGs seem to be associated with interactions$^{\cite{surace}}$.  
Occasionally quasars are also found to be associated
with old starburst populations 
($\sim 0.4$Gyr)$^{\cite{canalizo,brotherton}}$.  Simulations$^{\cite{mh}}$
show that gas can be driven into the inner regions of galaxies during
merger events and can trigger short-lived bursts of star-formation, and
this may also trigger accretion onto a massive
central black hole$^{\cite{barnes92}}$.  The most
luminous quasars ($M_R < -22$) are always found in massive host galaxies,
and it now appears that those host galaxies are elliptical galaxies
irrespective of whether the quasar is radio-loud or 
radio-quiet$^{\cite{mclure}}$:  
one route to the formation of massive elliptical galaxies
is through merging.
Overall, the hypothesis that emerges
is that quasars may be triggered by
interactions between galaxies and that the quasars are
relatively short-lived ($< 1 $Gyr) compared with the age of the universe.
In this case the dominant factor in the cosmological evolution of quasars
should be the rate at which galaxies merge.  
\cite{efrees} and \cite{haenhelt} have
previously invoked the merging of dark matter halos, and in this
paper we follow on from their work, 
but we present a new and more rigorous Press-Schechter$^{\cite{ps}}$
calculation of the merger rate between dark halos.  

\section{The Press-Schechter merger rate}

PS theory allows
us to estimate the number density of halos at any redshift as a function
of their mass$^{\cite{ps,bond}}$: 
halos of any given mass $M$ are being continually formed from mergers
between dark halos of lower mass.  Thus a calculation of the rate of
generation of halos of mass $M$ should give us an estimate of the
rate of merging at that halo mass.  However, it is not possible simply
to differentiate the PS mass-function wrt time because at any given cosmic
epoch not only are new halos continually being formed but also existing
halos of that mass are being lost into systems of yet higher mass.  
\cite{lc94}, {\em inter alia}, have considered conditional
probability distributions which allowed those authors to estimate
merger rates given a range of progenitor masses.  However, if we are
interested simply in the time evolution of the rate of mergers, we can
derive this much more simply from \cite{bond} as shown by
Percival (this volume) and \cite{wjplm}.  \cite{wjplm} argue that
the evolution in merger rate thus derived is independent of the
choice of progenitor masses. 
For a flat $\Omega_{\Lambda}=0$ cosmology 
this evolution is given by:
\begin{equation}
  \left(\frac{dn}{dt}\right)^{+}\propto(1+z)^{3.5}exp(-\beta(1+z)^{2})
  \label{eq:simple}
\end{equation} 
where $\beta=\alpha^{2}/(2\sigma_{M}^{2})$ is a function of the mass
of halo and the power spectrum.  The functional form is readily calculable
for other cosmologies$^{\cite{wjplm}}$.
In this paper we investigate four cosmologies (OCDM, $\Gamma$CDM,
SCDM, $\Lambda$CDM) (see \cite{wjplm}).

\section{Comparison with observations}

We need to convert the predicted rate of merging of dark matter
halos into observed quasar numbers.  We shall take a 
``black-box'' approach in which we ignore the unknown physics which
relates a merging dark halo to the luminosity of an activated quasar.
We shall instead assume the simplest possible model, namely that if
we take the rate of merging of dark halos in the mass range 
of the host galaxies of luminous quasars and convolve that
with an assumed quasar light-curve, then we shall obtain an estimate of the
observed number density of quasars at any epoch.  That is, we assume that
some fixed but unknown fraction of mergers  result in the activation
of a quasar. 
We need quasar lifetimes to be short compared with the Hubble time,
as otherwise the model would be convolved with a broad function
and would not match the large amount of observed evolution.  
In this study we simply choose an exponential
light curve of decay-time 0.6~Gyr.  

The OCDM model does not produce enough evolution. Reasonable fits can be
obtained for the remaining three models, although all of them have a
deficit in the amount of evolution between $z=2$ and $z=0$ of about a
factor 4.  We plot the $\Lambda$CDM model with halo mass 
$10^{11.8} \msun$ in Fig.~1. 
The normalisation has been treated as a free parameter
as the fraction of mergers (and indeed the actual typical
lifetime of quasars) is unknown. 
The SCDM model also fits reasonably with a high value for halo
mass, but can probably be excluded on other
grounds$^{\cite{gawiser}}$.  The $\Gamma$CDM model fits with a value
for the halo mass of $10^{10.6} \msun$: a value which is rather low,
although the halo
mass would be increased to $10^{11.8} \msun$ if $\sigma_8$ were
increased from 0.6 to 0.9.  
Increasing $h$ from 0.5 to 0.7 decreases the inferred halo masses by a
factor about 3.

\section{Discussion}

This picture produces good agreement with the key
features of the evolution of quasars:  a rapid rise in quasar space
density over $0 < z < 2$, a peak at $z \sim 2.5$ followed by a decline to
higher redshifts.  The halo mass inferred is comparable to that inferred
from the observation that quasars live in massive host galaxies.  The
models don't produce quite enough evolution over the range $0 < z < 2$
however:  a factor 4 more evolution appears to be required.  Previous
authors have suggested the expected cosmological evolution in either pairwise
velocity differences$^{\cite{carlberg}}$ or in halo rotation 
velocity$^{\cite{haenhelt}}$ may be a factor.  

It has also been suggested that the observed cosmological
evolution in star-formation rate and quasar numbers may be linked.  
If both are a by-product of mergers the model described here may
also apply to the star-formation
evolution.  Fig.~1 also shows the evolution in star-formation
history together with the predictions
of the merger model, again convolved with an exponential light curve of
decay-time 0.6~Gyr (appropriate for the ultraviolet light from a burst
of star formation$^{\cite{bruzual}}$).  We can see excellent agreement with the
data at $z < 2$.  Almost any halo mass gives the
required amount of evolution in the redshift range where the data are 
well-constrained.  There is thus effectively only one free parameter, the
normalisation of the curve, which is a measure of the efficiency
with which dark halo mergers generate stars.   Because the model assumes
that the star-formation occurs in short-lived bursts associated directly
with mergers we see a large amount of evolution, in contrast
to semi-analytic modelling$^{\cite{baugh}}$ where it would
appear that the smaller predicted amounts of evolution are primarily
due to having longer-lived periods of star formation.  A prediction
of the model presented here is that the stellar populations associated
with the evolving ultraviolet light should all appear young ($< 1 $Gyr).
At higher redshifts the data are still very uncertain, but we hope that
future determinations of the mass of the star-forming galaxies and their halos 
should be a good test of the model.

\end{document}